\documentstyle[multicol,prl,aps,epsfig]{revtex}
\newcommand{\etal}{{\it et al.}}

\newcommand{\MRp}{\mathrm{p}}

\newcommand{\gp}{g_1^{\mathrm{p}}}
\newcommand{\gfp}{g_1^{\mathrm{p}}/F_1^{\mathrm{p}}}
\newcommand{\pt}{P_{\mathrm{T}}}
\newcommand{\pta}{P_{\mathrm{T}}^{\mathrm{atom}}}
\newcommand{\ar}{\alpha_{\mathrm{r}}}

\begin{document}


\title{Measurement of the Proton Spin Structure Function ${\bf 
g_1^p }$ \\
  with a Pure Hydrogen  Target}


\author{
\centerline {\it The HERMES Collaboration}
A.~Airapetian$^{33}$, 
N.~Akopov$^{33}$, 
I.~Akushevich$^6$,
M.~Amarian$^{26,28}$, 
E.C.~Aschenauer$^{7}$, 
H.~Avakian$^{11}$, 
R.~Avakian$^{33}$, 
A.~Avetissian$^{33}$, 
B.~Bains$^{16}$, 
C.~Baumgarten$^{24}$
% ...
}

\address{
% ...
$^6$DESY, Deutsches Elektronen Synchrotron, 22603 Hamburg, Germany\\
$^7$DESY Zeuthen, 15738 Zeuthen, Germany\\
% ...
$^{11}$Istituto Nazionale di Fisica Nucleare, Laboratori Nazionali di Frascati, 00044 Frascati, Italy\\
% ...
$^{16}$Department of Physics, University of Illinois, Urbana, Illinois 61801, USA\\
% ...
$^{24}$Sektion Physik, Universit\"at M\"unchen, 85748 Garching, Germany\\
% ...
$^{26}$Nationaal Instituut voor Kernfysica en Hoge-Energiefysica (NIKHEF), 1009 DB Amsterdam, The Netherlands\\
% ...
$^{28}$Istituto Nazionale di Fisica Nucleare, Sezione Sanit\`a, 00161 Roma, Italy\\
% ...
$^{33}$Yerevan Physics Institute, 375036 Yerevan, Armenia
}


\maketitle

%
\begin{abstract}

A measurement 
of the  proton spin structure function 
$\gp(x,Q^2)$  
in deep-inelastic scattering is presented. 
The data were taken with the  27.6~GeV longitudinally polarised
  positron  beam at HERA incident on a longitudinally 
polarised pure hydrogen gas  
target internal to the storage ring.
The kinematic range is 
$0.021<x<0.85$ and
  0.8~GeV$^2<Q^2<20$~GeV$^2$.
The integral
  $\int_{0.021}^{0.85} \gp(x)\, dx$  evaluated at  $Q_0^2$ of 
  2.5~GeV$^2$ 
 is 
$0.122\pm 0.003$(stat.)$\pm 0.010$(syst.). 

\end{abstract}


\begin{multicols}{2}[]

Deep-inelastic lepton-nucleon  scattering 
 is well established 
as a powerful tool for the investigation  of  nucleon structure. 
The measured 
structure functions have been successfully interpreted
 in terms of 
 parton 
distributions.
Scattering   polarised leptons off polarised nucleons provides
information on the 
 spin 
composition of the nucleon.
The major aim of the HERMES experiment is to determine the spin 
contributions 
of the various  quark flavours 
to the 
spin of the nucleon 
from  the  combination of inclusive and 
semi-inclusive deep-inelastic polarised scattering data~\cite{us:proposal}.  

   At  centre-of-mass energies where weak contributions can be neglected, 
 inclusive polarised deep-inelastic  scattering is characterised
 by two spin structure functions: $g_1(x,Q^2)$ and $g_2(x,Q^2)$.
 Here   ${Q^2}$  is the negative  squared four-momentum 
of the exchanged
  virtual photon with   energy 
 $\nu$ and  {\small $x=Q^2/2M\nu$} 
is the Bjorken scaling
variable, where $M$ is the nucleon mass.  The fractional energy 
transferred to the nucleon is given  
by $y=\nu /E$ for  a lepton beam energy $E$.
 In leading order QCD the structure function $g_1$ is given by the 
  charge weighted sum
   over the 
 polarised 
quark (anti-quark) 
spin  distributions $\Delta q_f$ ($\Delta \overline {q}_f$):
\begin{equation}
\hspace*{-0.6cm} g_1(x,Q^2) = {1 \over 2}\sum_f e_f^2 \Big 
(\Delta q_f(x,Q^2)+\Delta \overline{q}_f(x,Q^2) \Big ).
\label{g1spin}
\end{equation}
Here $e_f$ is the charge of the quark (anti-quark) of flavour $f$ in 
units of the elementary charge 
and $x$ is interpreted as the fraction of the nucleon light-cone 
momentum carried by the struck quark.

The 
focus of this letter is 
the proton 
structure function $\gp$, which provides a powerful constraint
on the polarised quark distributions.
The  spin structure functions $g_1$ and $g_2$ 
 can be determined 
from   measurements of cross section asymmetries by combining 
data taken with a longitudinally polarised lepton beam and different  
spin orientations of the  target nucleons.
Scattering off 
 a longitudinally polarised target 
or a transversely 
polarised  target yields  
the asymmetries $A_{||}$ and $A_{\perp}$, respectively.
This letter reports on the 
results for 
 $\gp$ 
 from the measurement of  $A_{\parallel}$  
using inclusive deep-inelastic  scattering data  collected in 1997.
For this HERMES measurement the target 
was pure 
polarised hydrogen gas  without dilution by other unpolarised atomic 
species, in contrast to  
previous
 measurements of $\gp$ 
performed with solid targets
at SLAC~\cite{us:slac} and 
at  CERN~\cite{us:smc1,us:smc}.
The target thickness, fraction of polarisable material, and the polarimetry 
are all significantly different for solid and gaseous targets.

The HERMES experiment is
  located in the East straight section of the 
 HERA  storage ring at the DESY laboratory in Hamburg.
It uses the 
positron beam of 27.57 GeV energy with beam currents
decreasing 
typically from 40 to 10 mA in eight hours. The positrons become transversely
polarised by 
the emission of synchrotron radiation~\cite{us:soko}. 
Longitudinal polarisation of the positron beam at the interaction
point is achieved with spin rotators~\cite{us:barber} situated 
upstream and downstream of the HERMES experiment.
Equilibrium 
 polarisation values in the range of 0.40 
to 0.65 
are  reached  with a rise-time  of about 30 minutes. 
The beam polarisation is continuously measured 
using
 Compton back-scattering of circularly polarised laser light, achieving a 
  statistical accuracy  of typically 1\% in 60~s.
Two 
 polarimeters are used, one measuring the transverse polarisation in the HERA 
 West straight section~\cite{us:TPOL} 
and the other  measuring the longitudinal 
polarisation 
 near the HERMES target~\cite{us:most}.
 After a normalisation by rise-time
measurements they give consistent results 
for the entire running period~\cite{us:lorenzon}. 
The systematic uncertainties are 
respectively 3.4\% and
4.3\% of the measured values, dominated by the normalisation uncertainty of
3.3\% as determined from the rise-time calibration.
Data were accepted for this $\gp$ analysis  when the 
 polarisation value  was above 0.30; 
the average  polarisation
 for this data was 0.55. 
For about 12\% of the data,
  only the longitudinal polarimeter was operational.

The HERMES polarised proton target~\cite{us:target,us:braun}   is formed
by
injecting a nuclear-polarised beam  of  atomic hydrogen
  from an atomic beam source (ABS)~\cite{us:stock} into a 
tubular open-ended storage cell.
The cell  confines the atoms 
in the region of the circulating beam 
 and increases  the
probability of a positron-proton interaction by a factor of approximately
one hundred compared with the free atomic beam.
The  storage cell 
 is 
made of 75~$\mu$m thick aluminium, 400~mm long, and has an elliptical
cross section 9.8~mm high and 29~mm wide.  It 
is cooled to 100~K giving an
areal density of $7 \times 10^{13}$
atoms/cm$^2$.
 The proton polarisation in 
the atomic beam is  above  0.95, while the electron
polarisation is less than 0.05. The beam also contains unpolarised
hydrogen molecules at a level  less than 1\%.  
Due to depolarisation mechanisms which take
place within the cell, the proton polarisation within the 
 cell 
is slightly  
lower than in the atomic beam. These are spin exchange in 
atom-atom collisions, as well as 
depolarisation and recombination to molecules
 during interactions with the cell walls. 
The cell walls are coated with drifilm~\cite{us:dri}
to minimise the wall collision effects. There is good evidence that
recombination is further suppressed by water adsorbed on the cell
surface during normal operation~\cite{us:target,us:hauke}. A static
magnetic field directed parallel to the positron beam axis is provided
throughout the cell to define the quantisation axis.
The operational field value is chosen to avoid resonant depolarisation of
the protons by the pulsed magnetic field caused by the bunch
structure of the  HERA beam~\cite{us:depolpaper}. The spin direction in 
the target can be
reversed in less 
than one second by selecting different spin states in the ABS. In operation
the length of the time interval between reversals was randomised and was
of the order of one minute. The atomic beam is injected into the storage
cell via a side tube connected to its centre. The atoms
diffuse to 
the open ends of the cell generating a triangular
density distribution along the beam axis.  The escaping gas is removed
from the storage ring by high speed vacuum pumps. The residual gas
in this vacuum system produces a further small source of unpolarised
hydrogen molecules in the storage cell. The gas at the centre of the cell is
sampled via a second side tube.  Both the nuclear and electron 
polarisations of the atoms in this
sample are  
measured with a Breit-Rabi polarimeter (BRP), and the atomic fraction with a
target gas analyser (TGA)~\cite{us:braun}. These measurements are
made continuously 
achieving a 
 statistical accuracy of typically 0.02 in  60~s.

The target proton polarisation  $\pt$  as seen by the positron beam is given by
\begin{equation}
\pt=\alpha_0\left[ \ar+(1-\ar)\beta\right] \pta.
\label{targetpol}
\end{equation}  
Here $\pta$ is the proton polarisation in the atoms; its value was 
0.92 $\pm$ 0.03. The atomic fraction $\alpha_0$ accounts for the presence of
the  small number of  molecules originating from the ABS and the 
residual gas in the 
vacuum system; it had a  value of 0.99  $\pm$ 0.01. The fraction of
 atoms that are in the form of molecules produced by recombination is
$(1 - \ar)$.  The  value of $\ar$ was 0.93 $\pm$ 0.04. 
The values of $\pta$ and $\ar$ were corrected for the sampling 
efficiencies of the BRP and TGA. These corrections  depend on the 
measured values 
and the knowledge of depolarisation mechanisms  inside the cell. 
The quantity   $\beta$ is defined
as the ratio of the polarisation of protons in molecules from
recombination to the polarisation of protons in the atoms. 
Limits on this ratio ($0.2 \leq \beta \leq 1.0$) were derived~\cite{us:hauke} 
from the relationship  between the target polarisation and the corresponding
asymmetry $A_{||}$ 
 measured when the  recombination rate was high 
 due to deliberately modified conditions at the cell surface.
The resulting target polarisation $\pt$ was 0.88 $\pm$ 0.04. 
The quoted values are
averages over the data taking period and 
the uncertainties  are systematic. 

The HERMES detector is a 
forward spectrometer 
with a  dipole magnet providing an integrated field of 1.3~Tm. 
The magnet 
 is divided into two identical sections 
by a horizontal iron plate that shields the 
positron and
proton beams 
from the
magnetic
field.
Consequently, the spectrometer consists of two identical detector systems,
and  the minimum polar angle for the  acceptance of    scattered
positrons  is 40~mrad.  
The maximum angular acceptances are $\pm$140~mrad vertically and 
$\pm$170~mrad horizontally.
 For tracking in each spectrometer half,  42 
drift chamber
planes  
and 6 micro-strip gas chamber planes are used. Fast track 
reconstruction is achieved by a pattern-matching
algorithm and a momentum look-up method~\cite{us:wwc}. For
 positrons  with  momenta between  3.5 and 27~GeV,
the  average  
angular resolution is 0.6-0.3~mrad and the 
average momentum resolution  $\Delta p/p$ is 0.7-1.3\% 
aside from bremsstrahlung tails.
A detailed description of the spectrometer is found 
in Ref.~\cite{us:detector}.

The trigger is formed 
by a coincidence of signals from  three hodoscope
planes with those  
from a lead-glass calorimeter,
requiring
 an energy of greater than 1.5~GeV to be deposited locally in the  calorimeter.
Positron
identification  is accomplished 
 using the calorimeter, a scintillator hodoscope 
preceded by two radiation lengths of 
lead, 
a six-module transition radiation detector, 
and a C$_4$F$_{10}/$N$_{2}$(70:30)
 gas threshold 
 \v{C}erenkov
counter.

 An adjustable two-stage collimator system is  mounted upstream of
the target cell to  protect
target and spectrometer  from  synchrotron radiation and from beam halo. 
 The number of triggers originating from particles scattered from the 
storage cell walls was 
 negligible.
The  luminosity is measured by detecting  electron-positron pairs
from Bhabha scattering off the target gas  electrons, in two 
 NaBi(WO$_4$)$_2$ electromagnetic calorimeters, which are mounted symmetrically
on either side of the beam line. 

The  cross section asymmetry  $A_{||}$ 
was 
determined 
using the formula 
\begin{equation}
{A_{||} } = {N^-L^+ - N^+L^- \over N^-L^+_{\rm{P}} + N^+L^-_{\rm{P}}}.
\label{eq:ap}
\end{equation}
Here $N^+ (N^-)$ is the number of scattered positrons for target spin parallel
(anti-parallel) to the beam spin orientation. 
 The 
deadtime corrected
luminosities for each target spin state
are  $L^{\pm}$ and 
$L^{\pm}_{\rm{P}}$, 
  the latter  being weighted by  the product of the
 beam and target polarisation values for each spin state.
The 
luminosities  
were corrected for a small Bhabha cross section 
asymmetry 
 caused by the typically 3\% polarisation  of the 
 electrons in the target. The effect on $A_{||}$ 
was determined to better than 0.2\%.

The 
kinematic requirements imposed on the data were: $Q^2 > 0.8$
GeV$^2$, $0.1 <y < 0.85$,  an energy of the hadronic final 
state $W> 1.8$ GeV, and  a
minimum 
calorimeter energy deposition of 3.5 GeV. 
After applying data quality criteria, $1.7\times 10^6$
events were available for the asymmetry analysis.  
The data were analysed in bins of  $x$ and also $y$ 
 in order to include the effect of the  variation of 
the virtual photon depolarisation factor $D$ with $y$. 
For each bin and spin state the 
number of scattered positrons  was corrected for $e^+e^-$ 
background from charge symmetric processes. 
This correction was  at most 8\% in the bins with   $x
<0.08$ and $y>0.6$, and  negligible  for the remainder of the kinematic range. 
For the average positron identification efficiency of 99\%,
 the average 
contamination of
misidentified hadrons was negligible
 with values not greater than 1\% in 
the  
lowest $x$ bins. 
 The asymmetry was  further corrected for
 smearing
effects due to the finite resolution of the
spectrometer, which 
were determined  by Monte Carlo simulations to be about 8\% at low $x$ 
and in the range of 2-3\% 
at high $x$.
QED radiative correction factors  were
calculated following the prescription given in Ref.~\cite{radcor}.
The corrections were determined to 
be less than  8\% at low $x$ and high $y$, decreasing to 
 less than 1\% at  higher $x$ values.
Both smearing  and radiative corrections 
were  calculated 
using  an iterative procedure.

The structure function ratio $g_1/F_1$ is 
approximately equal to the longitudinal virtual photon asymmetry $A_1$.
 It was calculated in each ($x,\,y$) bin from the 
longitudinal  asymmetry $A_{||}$, 
corrected for the effects mentioned above, 
using the relation 
\begin{equation}
{g_1\over F_1} = {1 \over 1+\gamma^2}\left[ {A_{||}\over D}  + 
(\gamma- \eta )
  A_2\right].\label{g1eq}
\end{equation}
The virtual photon depolarisation
factor
$D =[1-(1-y)\epsilon]/(1+\epsilon R)$  depends on the ratio
 {\small $R=\sigma_L/\sigma_T$} of longitudinal to
transverse virtual  photon absorption cross sections and 
is approximately  equal to $y$. 
 The kinematic factors 
 are defined as  
$\epsilon = [4(1-y) - \gamma^2 y^2]/[2y^2+4(1-y)+\gamma^2 y^2]$, 
 $\gamma=2Mx/\sqrt{Q^2}$, and 
$\eta = \epsilon \gamma y/[1-\epsilon(1-y)]$. 
The magnitude of the  transverse virtual photon absorption 
asymmetry $A_2^{\MRp}$ 
has been measured
previously to be small~\cite{us:slac}.
Its contribution to $\gfp$  is further 
suppressed by the factor ($\gamma-\eta$) 
  and was taken into account  using  
a fit  based on   existing data~\cite{us:slac,us:smc}:  
$A_2^{\MRp}=0.5\cdot x/ \sqrt{Q^2}$. 

\begin{figure}[htb]
\begin{center}
\vspace*{-0.5cm}
\epsfxsize 9 cm {\epsfbox{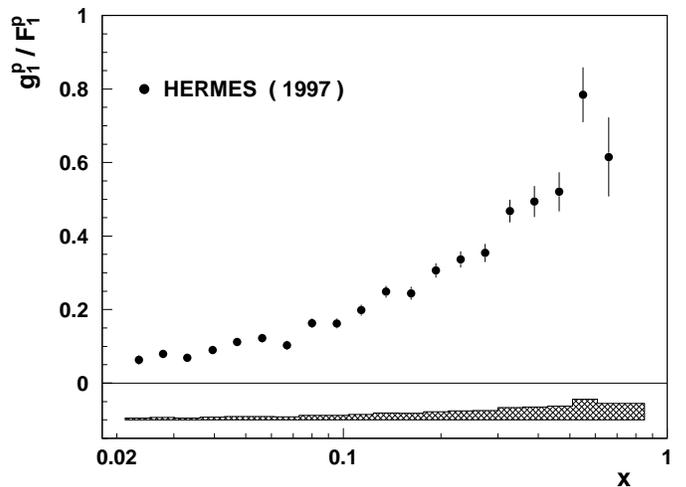}}
\begin{minipage}[r]{\linewidth}
\caption{
The  structure function ratio  $\gfp$ of the 
 proton 
as a function of $x$, given for the  measured $\left<Q^2\right>$ at 
each value of $x$.
  The  error 
bars show the statistical uncertainties  and the  band represents
the total systematic uncertainties.}
 \label{a1plot1}
\end{minipage}
\end{center}
\end{figure}

\begin{figure}[h]
\begin{center}
\vspace*{-0.7cm}
\epsfxsize 18 cm {\epsfbox{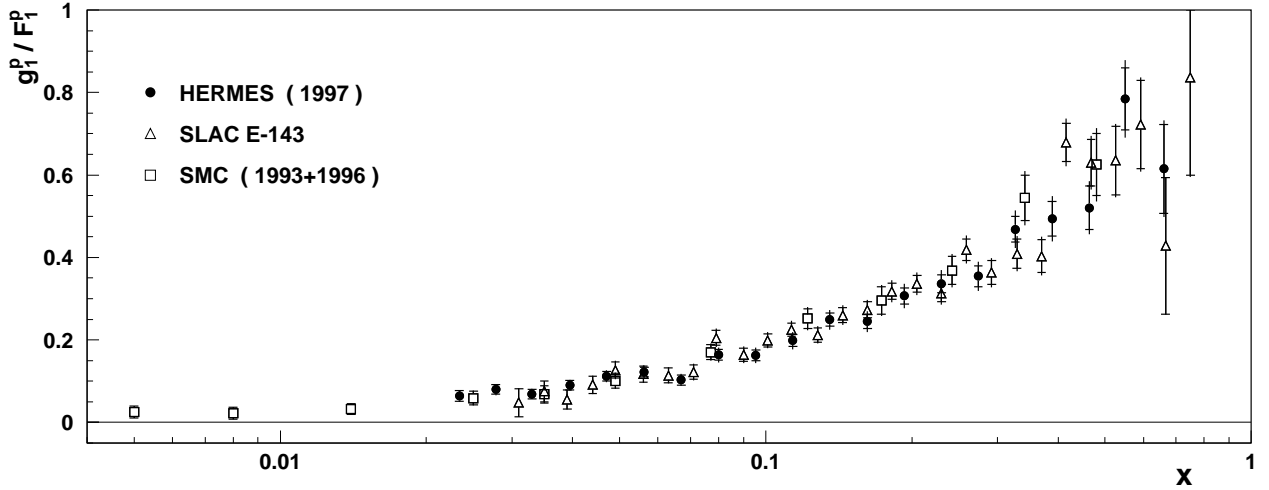}}
\begin{minipage}[r]{\linewidth}
\caption{
Comparison of the data shown in Fig.~\ref{a1plot1} with recent 
 results for $\gfp$ obtained at
  SLAC (E-143) and  $A_1^{\MRp}$  at CERN (SMC) for $Q^2>1$~GeV$^2$.
 The inner error bars show the
  statistical uncertainties and  the outer ones the quadratic sum of 
  statistical  and total systematic uncertainties.}\label{a1plot2}
\end{minipage}
\end{center}
\end{figure}

\begin{figure}[h]
\begin{center}
\vspace*{-0.7cm}
\epsfxsize 18 cm {\epsfbox{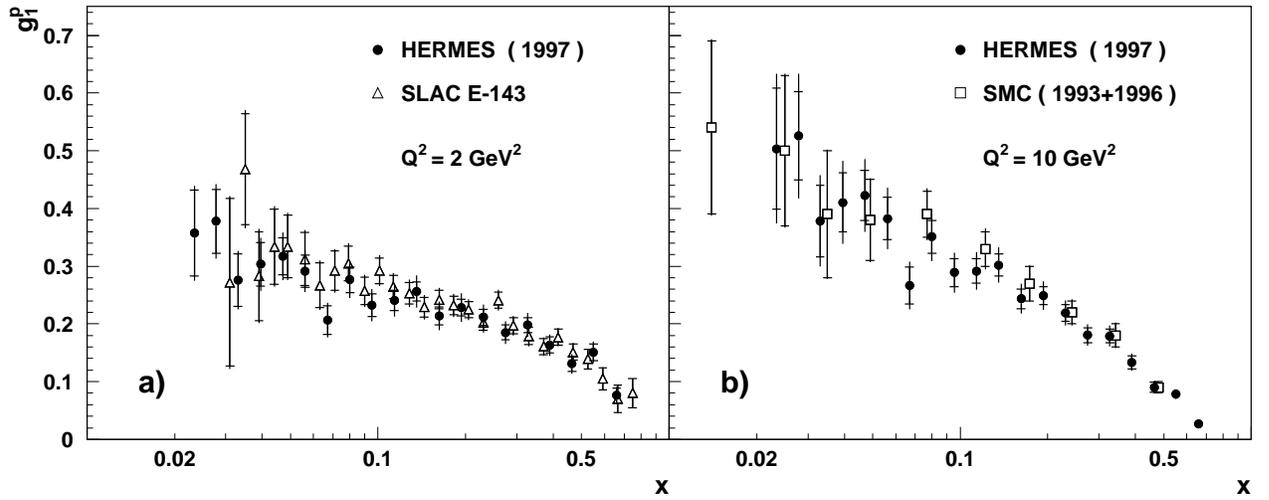}}
\begin{minipage}[r]{\linewidth}
\caption{
The  spin structure function $\gp$ of the proton as a
function of $x$. The HERMES data are evolved to $\textrm{a})$
$Q_0^2=2$~GeV$^2$  and $\textrm{b})$ $Q_0^2=10$~GeV$^2$
assuming $\gfp$ to be independent of $Q^2$. This measurement is 
compared to recent results for $Q^2>1$~GeV$^2$ from 
E-143~\protect\cite{us:slac}
and from SMC~\protect\cite{us:smc}, 
the latter shown for $x>0.01$ only. The error 
bars are defined as in Fig.~\ref{a1plot2}.
}
\label{g1plot}
\end{minipage}
\end{center}
\end{figure}

The $x$ dependence of the  structure function ratio 
$\gfp$ 
at the measured   $\langle Q^{2}\rangle$ for each value of $x$  is shown in 
Fig.~\ref{a1plot1}.
The averaged 
values of $x$,  $Q^2$, 
 and the structure function ratio $\gfp$ are listed in Table~\ref{g1table}.
Fig.~\ref{a1plot2} 
shows
a comparison of the measured  $\gfp$ values with recent 
results of previous experiments,
E-143 ~\cite{us:slac}
and SMC~\cite{us:smc}. There is good agreement between 
these three sets of data, 
although the  $\langle Q^2 \rangle$ values of E-143 and HERMES  
differ from those of SMC by a 
factor  between five and ten. This demonstrates that there is no 
statistically significant $Q^2$
dependence of the ratio $\gfp$ in the $Q^{2}$ range of these 
experiments. 
The spin structure function $\gp(x,Q^2)$ was  extracted
from the ratio $\gfp$ 
using the relation  
$F_1 = F_2 (1+\gamma^2)/(2x(1+R))$  with parameterisations of
the unpolarised structure function $F_2^{\MRp}(x,Q^2)$~\cite{us:nmc} 
and $R(x,Q^2)$~\cite{us:R}.
The  values of 
$\gp$  
at the measured $x$ and $Q^2$ and after evolution to a common $Q_0^2$ 
of 2.5~GeV$^2$ are also given in Table~\ref{g1table}. 
The evolution was done under the assumption that the 
ratio of $\gfp$ does not depend on $Q^2$.
The same assumption was used to evolve the HERMES data to the 
$Q_0^2$ values as published by E-143 and SMC. 
As shown in 
 Figs.~\ref{g1plot}a and ~\ref{g1plot}b,
 the evolved HERMES data  
are in excellent agreement with both data sets.
The apparent $Q^2$ dependence of $\gp$ seen 
 when comparing Fig.~\ref{g1plot}a with   Fig.~\ref{g1plot}b entirely
 originates from the $Q^2$ dependence of the 
unpolarised structure functions $F_2^{\MRp}$ and $R$.

The systematic uncertainty of this   measurement of $\gfp$, 
illustrated 
by the  band in
Fig.~\ref{a1plot1}, is about 8\% over the entire $x$ range.
The dominant sources of
systematic uncertainties  are 
the beam and target polarisation  discussed above. 
An uncertainty of 2.5\%  is included to account for the 
uncertainty in  the description of the spectrometer geometry.  
The uncertainty  originating from  the combined effects of  
smearing and QED radiative corrections is between 1\% and 4\%. 
The data were 
 searched for possible systematic  fluctuations in the measured asymmetry.
This was carried out  by dividing the data   
into 
sub-samples defined by  various parameters, {\em e.g.}  time, detector 
geometry and beam current. 
Within the statistical accuracy of the data  no 
additional systematic effect was detected. 
\end{multicols}


\begin{multicols}{2}[] 

The effect of the assumption for $A_{2}^{\MRp}$ on the extracted ratio 
$\gfp$ was estimated from an analysis of
  the variation within the  statistical and systematic 
 uncertainties  of the 
 recent $A_2^{\MRp}$ data of E-143~\cite{us:slac},
  the assumption $g_2=0$, and the use of the  ansatz for 
 $g_2$ given in Ref.~\cite{us:ww}. 
The systematic 
uncertainty due to this effect 
was estimated to be at most 1.5\%. 

The values  of the  ratio $\gfp$ and also $\gp$ derived from 
$A_{||}$
depend on $R$. 
Using the uncertainties for $R$ given in Ref.~\cite{us:R}  changed   
the  $\gfp$ ($\gp$) values by typically 3\% (2\%) and 
at most 5.5\% (3.6\%) at $x<0.1$.
In the HERMES kinematic region this uncertainty range of $R$ covers the 
$R$ measurement recently published for $x<0.12$~\cite{us:RNMC}. 
Variations in the  $F_2^{\MRp}$ 
parameterisation calculated in Ref.~\cite{us:nmc} 
including the  normalisation 
uncertainty of the $F_2^{\MRp}$ values of 0.7\%  resulted in   changes in the 
extracted $\gp$  of about   2.5\%. 
The influence of $R$ and $F_{2}^{\MRp}$ has not been included in the 
systematic uncertainty of the measurement and  the values are given separately 
for each $x$ bin in Table~\ref{g1table}.
\end{multicols}
\begin{table}[t]
\caption{ Results on $\gp(x,Q^{2})/F_1^{\MRp}(x,Q^{2})$,
$\gp(x,Q^{2})$  
 and $\gp(x,Q_0^2)$ evolved to $Q_0^2=2.5~$GeV$^2$ assuming $\gfp$ to be  independent of $Q^2$. 
The measured  $\left<Q^2\right>$ values are given in GeV$^2$.
 For each $x$ bin the uncertainties concerning $R$ and $F_2$,  $\delta(R)$ or $\delta(R,F_2)$, are listed separately. 
The systematic uncertainties are dominated by those of the beam and target polarisation which constitute a normalisation uncertainty only.}
\label{g1table} \medskip
\begin{tabular}{ccccc} 
 $\left<x\right>$ & $\left<Q^2\right>$  &    $\gfp \pm$stat.$\pm$syst.$\pm \delta(R)$&$\gp \pm$stat.$\pm$syst.$\pm \delta(R,F_2)$&$\gp$($x,Q_0^2$)$\pm$stat.$\pm$syst.$\pm \delta(R,F_2)$  
\\
\hline
 0.023 & 0.92 & 0.064 $\pm$ 0.013 $\pm$ 0.004 $\pm$ 0.002 & 0.300 $\pm$ 0.062 $\pm
$ 0.021 $\pm$ 0.012 & 0.375 $\pm$ 0.078 $\pm$ 0.035 $\pm$  0.017 \\ 
 0.028 & 1.01 & 0.080 $\pm$ 0.012 $\pm$ 0.006 $\pm$ 0.003 & 0.327 $\pm$ 0.048 $\pm
$ 0.023 $\pm$ 0.012 & 0.395 $\pm$ 0.058 $\pm$ 0.034 $\pm$  0.016 \\ 
 0.033 & 1.11 & 0.069 $\pm$ 0.011 $\pm$ 0.005 $\pm$ 0.003 & 0.245 $\pm$ 0.041 $\pm
$ 0.017 $\pm$ 0.008 & 0.288 $\pm$ 0.047 $\pm$ 0.024 $\pm$  0.011 \\ 
 0.040 & 1.24 & 0.090 $\pm$ 0.011 $\pm$ 0.006 $\pm$ 0.004 & 0.284 $\pm$ 0.035 $\pm
$ 0.019 $\pm$ 0.007 & 0.316 $\pm$ 0.039 $\pm$ 0.029 $\pm$  0.010 \\ 
 0.047 & 1.39 & 0.112 $\pm$ 0.011 $\pm$ 0.007 $\pm$ 0.004 & 0.302 $\pm$ 0.031 $\pm
$ 0.020 $\pm$ 0.008 & 0.329 $\pm$ 0.034 $\pm$ 0.029 $\pm$  0.010 \\ 
 0.056 & 1.56 & 0.122 $\pm$ 0.012 $\pm$ 0.008 $\pm$ 0.004 & 0.284 $\pm$ 0.027 $\pm
$ 0.019 $\pm$ 0.007 & 0.302 $\pm$ 0.029 $\pm$ 0.025 $\pm$  0.008 \\ 
 0.067 & 1.73 & 0.103 $\pm$ 0.012 $\pm$ 0.007 $\pm$ 0.004 & 0.206 $\pm$ 0.025 $\pm
$ 0.014 $\pm$ 0.005 & 0.213 $\pm$ 0.026 $\pm$ 0.017 $\pm$  0.006 \\ 
 0.080 & 1.90 & 0.163 $\pm$ 0.013 $\pm$ 0.011 $\pm$ 0.006 & 0.280 $\pm$ 0.022 $\pm
$ 0.018 $\pm$ 0.008 & 0.285 $\pm$ 0.023 $\pm$ 0.021 $\pm$  0.008 \\ 
 0.095 & 2.09 & 0.163 $\pm$ 0.014 $\pm$ 0.011 $\pm$ 0.006 & 0.240 $\pm$ 0.020 $\pm
$ 0.016 $\pm$ 0.007 & 0.238 $\pm$ 0.020 $\pm$ 0.018 $\pm$  0.007 \\ 
 0.114 & 2.26 & 0.198 $\pm$ 0.015 $\pm$ 0.013 $\pm$ 0.007 & 0.251 $\pm$ 0.019 $\pm
$ 0.017 $\pm$ 0.007 & 0.246 $\pm$ 0.018 $\pm$ 0.019 $\pm$  0.007 \\ 
 0.136 & 2.44 & 0.249 $\pm$ 0.016 $\pm$ 0.016 $\pm$ 0.009 & 0.268 $\pm$ 0.017 $\pm
$ 0.018 $\pm$ 0.008 & 0.261 $\pm$ 0.017 $\pm$ 0.020 $\pm$  0.007 \\ 
 0.162 & 2.63 & 0.244 $\pm$ 0.017 $\pm$ 0.016 $\pm$ 0.008 & 0.225 $\pm$ 0.016 $\pm
$ 0.015 $\pm$ 0.006 & 0.217 $\pm$ 0.015 $\pm$ 0.017 $\pm$  0.006 \\ 
 0.193 & 2.81 & 0.307 $\pm$ 0.019 $\pm$ 0.020 $\pm$ 0.010 & 0.237 $\pm$ 0.015 $\pm
$ 0.016 $\pm$ 0.007 & 0.231 $\pm$ 0.015 $\pm$ 0.017 $\pm$  0.007 \\ 
 0.230 & 3.02 & 0.336 $\pm$ 0.022 $\pm$ 0.022 $\pm$ 0.010 & 0.216 $\pm$ 0.014 $\pm
$ 0.014 $\pm$ 0.006 & 0.212 $\pm$ 0.014 $\pm$ 0.015 $\pm$  0.006 \\ 
 0.274 & 3.35 & 0.354 $\pm$ 0.025 $\pm$ 0.023 $\pm$ 0.011 & 0.183 $\pm$ 0.013 $\pm
$ 0.012 $\pm$ 0.005 & 0.184 $\pm$ 0.013 $\pm$ 0.013 $\pm$  0.005 \\ 
 0.327 & 3.76 & 0.468 $\pm$ 0.031 $\pm$ 0.031 $\pm$ 0.014 & 0.187 $\pm$ 0.012 $\pm
$ 0.012 $\pm$ 0.005 & 0.195 $\pm$ 0.013 $\pm$ 0.014 $\pm$  0.005 \\ 
 0.389 & 4.25 & 0.494 $\pm$ 0.042 $\pm$ 0.032 $\pm$ 0.015 & 0.147 $\pm$ 0.013 $\pm
$ 0.010 $\pm$ 0.004 & 0.158 $\pm$ 0.013 $\pm$ 0.011 $\pm$  0.004 \\ 
 0.464 & 4.80 & 0.520 $\pm$ 0.053 $\pm$ 0.034 $\pm$ 0.016 & 0.104 $\pm$ 0.011 $\pm
$ 0.007 $\pm$ 0.003 & 0.122 $\pm$ 0.013 $\pm$ 0.008 $\pm$  0.003 \\ 
 0.550 & 5.51 & 0.784 $\pm$ 0.075 $\pm$ 0.051 $\pm$ 0.024 & 0.094 $\pm$ 0.009 $\pm
$ 0.006 $\pm$ 0.003 & 0.134 $\pm$ 0.013 $\pm$ 0.009 $\pm$  0.003 \\ 
 0.660 & 7.36 & 0.615 $\pm$ 0.108 $\pm$ 0.040 $\pm$ 0.020 & 0.032 $\pm$ 0.005 $\pm
$ 0.002 $\pm$ 0.001 & 0.068 $\pm$ 0.012 $\pm$ 0.005 $\pm$  0.001 \\ 
\end{tabular}
\end{table}

\begin{multicols}{2}[]
Nucleon spin structure functions may be characterised in terms of sum rules 
which involve  integrals of $g_1$ over $x$ for a given $Q_0^2$~\cite{us:bj}.
The integral of $\gp(x)$  evaluated in the measured 
region $0.021<x<0.85$  and  $Q^2>0.8$~GeV$^2$
is   $0.122\pm 0.003$(stat.)$\pm 0.010$(syst.) at $Q_0^2$ of 2.5~GeV$^2$.
In the integration the $x$ dependence of $F_1^{\MRp}$ within the 
individual $x$ bins was
fully accounted for, whereas the
$x$ dependence of $\gfp$ was 
treated in first order only.
Possible $Q^2$ dependences not excluded by the present statistical accuracy
were investigated using $Q^2$ dependent fits to the $\gfp$ data 
according to Ref.~\cite{us:slac} and  a $Q^2$ dependence of $A_1^{\MRp}$ as  
parameterised by a next-to-leading order QCD analysis~\cite{us:NLO2}. 
The effect on $\gp$ was  at most  8\% at the lowest $x$ and $Q^2$ values 
and was included in the systematic uncertainty of the integral.
A 
contribution to the  systematic uncertainty  of 0.003 
arose from the uncertainties 
due to  $R$ and $F_2^{\MRp}$  described above.
The result obtained for the integral was compared with those 
from  E-143~\cite{us:slac} and SMC~\cite{us:smc}, both calculated
with the same integration scheme and for the kinematic range of HERMES.
With respect to statistical uncertainties the agreement 
is better than 0.6$\sigma$ and 1.2$\sigma$, respectively, well inside 
the normalisation uncertainty given by each experiment.

In summary, the proton structure function ratio $\gfp$ was 
measured  with good statistical and systematic precision.
The results were obtained using an entirely different technique from 
that  used in all previous experiments, involving a longitudinally 
polarised positron beam in a high energy storage ring and an internal
polarised pure  hydrogen gas target.
The data are in good agreement with
those obtained with solid targets at both similar and much higher
$Q^2$ values, indicating that the systematic uncertainties are well
understood for both techniques over the entire measured $x$ range.
Evolved to the same values of  $Q^2$ all recent data on $\gp$ and on its
integral are consistent.

We gratefully acknowledge the DESY management for its support and 
the DESY staff and the staffs of the collaborating institutions. 
This work was supported by  
the FWO-Flanders, Belgium;
the Natural Sciences and Engineering Research Council of Canada;
the INTAS, HCM, and TMR network contributions from the European Community;
the German Bundesministerium f\"ur Bildung, Wissenschaft, Forschung
und Technologie; the Deutscher Akademischer Austauschdienst (DAAD);
the Italian Istituto Nazionale di Fisica Nucleare (INFN);
Monbusho, JSPS, and Toray
Science Foundation of Japan;
the Dutch Foundation for Fundamenteel Onderzoek der Materie (FOM);
the U.K. Particle Physics and Astronomy Research Council; and
the U.S. Department of Energy and National Science Foundation.


\end{multicols}

\end{document}